\documentclass[review,square,sort&compress]{elsarticle}

\usepackage{graphicx}
\graphicspath{{figures/}}
\DeclareGraphicsExtensions{.pdf,.jpg,.png}

\usepackage{amssymb}
\journal{Nucl. Instr. and Meth. in Phys. Res. A}

\def\KAOS{{\sc Kaos}\@}
\def\PANDA{$\overline{\mbox{P}}${ANDA}\@}

\begin{document}

\begin{frontmatter}
\title{Future use of silicon photomultipliers for {\sc Kaos}
  at MAMI and \PANDA\ at FAIR} 

\author[a]{P.~Achenbach\corref{pa}}
	\ead{patrick@kph.uni-mainz.de}
\author[a]{A.~{Sanchez Lorente}}
\author[a]{S.~{S\'anchez Majos}}
\author[a]{J.~Pochodzalla}

\cortext[pa]{Corresponding author. Tel.: +49-6131-3925831;
  fax: +49-6131-3922964.}

\address[a]{Institut f\"ur Kernphysik, Johannes Gutenberg-Universit\"at
    Mainz, Germany}

\begin{abstract}
  A characterisation of scintillating fibres with silicon
  photomultiplier read-out was performed in view of their possible
  application in fibre tracking detector systems.  Such a concept is
  being considered for the \KAOS\ spectrometer at the Mainz Microtron
  MAMI and as a time-of-flight start detector for the hypernuclear
  physics programme at the \PANDA\ experiment of the FAIR project.
  Results on particle detection efficiency and time resolution are
  discussed. In summary, the silicon devices are very suitable for the
  detection of the low light yield from scintillating fibres insofar a
  trigger scheme is found to cope with the noise rate characteristics.
\end{abstract}


\end{frontmatter}

\section{Introduction}
The time-of-flight (TOF) method is often being used for particle
identification in the low momentum range ($p <$ 1\,GeV$/c$) of
magnetic spectrometers. A typical detector package in such a system
needs to provide TOF, specific energy loss (dE$/$dx) and track
information and consists of fast scintillators situated behind wire
chambers.  Alternatively, a detector based on small cross-section
scintillating fibres can combine the TOF and dE$/$dx information with
coordinate information. Several requirements must be fulfilled of such
a system to be competitive to more conventional solutions:
\begin{enumerate}
\item high particle detection efficiency ($\epsilon \simeq$ 100\,\%)
  at low mass ($x \leq$ 5\,g$/$cm$^2$),
\item fast time response (FWHM $<$ 1\,ns) and high counting rate
  capability,
\item sufficient stability of the detector response in a demanding
  environment, i.e.\ under strong irradiation, inside a varying
  magnetic field, operation in vacuum.
\end{enumerate}

The silicon photomultiplier (SiPM) is a novel semiconductor
photodetector operated in the limited Geiger mode, capable of
resolving individual photons~\cite{Renker2006}. In combination with
scintillating fibres a SiPM can provide a relatively cheap and
reliable tracking detector operated with low voltage, magnetic field
insensitive and minimal volume~\cite{Dolgoshein2006}. The SiPM can
have advantages over other read-out devices like multi-anode
photomultipliers, e.g.\ when power dissipation is a critical issue
(inside the vacuum system of the \KAOS\ spectrometer at the Mainz
Microtron MAMI), or when space restrictions are tight (close to the
planned hypernuclear target at the \PANDA\ experiment).

For this paper the characteristics of SiPM from different
manufacturers coupled to round and square scintillating fibres of
several diametres were studied in view of their possible application
in these two experiments.

\section{SiPM in {\sc Kaos} tracking detectors}
At the Institut f\"ur Kernphysik in Mainz, Germany, the microtron MAMI
has been upgraded to 1.5\,GeV electron beam energy and can now be used
to study strange hadronic systems. A large fibre detector set-up is
under development for the \KAOS\ spectrometer~\cite{Achenbach-SNIC06}
of the A1 Collaboration: the coordinate detector of the spectrometer's
electron arm will consist of two planes of vertical fibre arrays,
covering an active area of $L \times H\sim$ 2000\,mm $\times$ 300\,mm,
supplemented by one or more planes with horizontal fibres. The
read-out of the vertical fibres will be performed one-sided by
multi-anode photomultipliers. The experience acquired in the design
and beam-test of these detectors has suggested that the use of
multi-anode photomultipliers with several fibres per pixel introduces
a relatively high degree of complexity. SiPM have been suggested as a
possible candidate for a two-sided read-out for the long fibres in
horizontal direction. Their use would simplify detector mechanics and
reduce considerably the over-all cost.

Up to now the high intrinsic noise count rate of SiPM has limited
their use to cases in which tens or hundreds of photons are available,
but low light yields have been measured with 0.83\,mm diametre,
cylindrical fibres. Gain variations from diode to diode and
temperature dependence are further problems to be considered.

\begin{table}[htbp]
  \centering
  \caption{Particle detection efficiencies as a function of 
    discriminator threshold in units of single pixel amplitude 
    as measured with a three reference counter system. Set-up~A 
    consisted of a 2\,m long round fibre of 0.83\,mm diametre 
    with 1\,mm$^2$ SiPM read-out and set-up~B of a 2\,m long 
    square fibre of 2\,mm width with 4.4\,mm$^2$ SiPM read-out.}

  \begin{tabular}{lll}
    \hline
    Threshold        & Set-up A        & Set-up B\\
    (no.\ of pixels) & Efficiency (\%) & Efficiency (\%)\\

    \hline
    \hline
    0.5 & 91 & 100\\
    1.5 & 76 & 99.8\\
    2.5 & 56 & 95.0\\
    3.5 & 35 & 82.6\\
    \hline
  \end{tabular}
  \label{tab:eff}
\end{table}
\begin{figure}[htbp]
  \centering
  \includegraphics[width=0.7\textwidth]{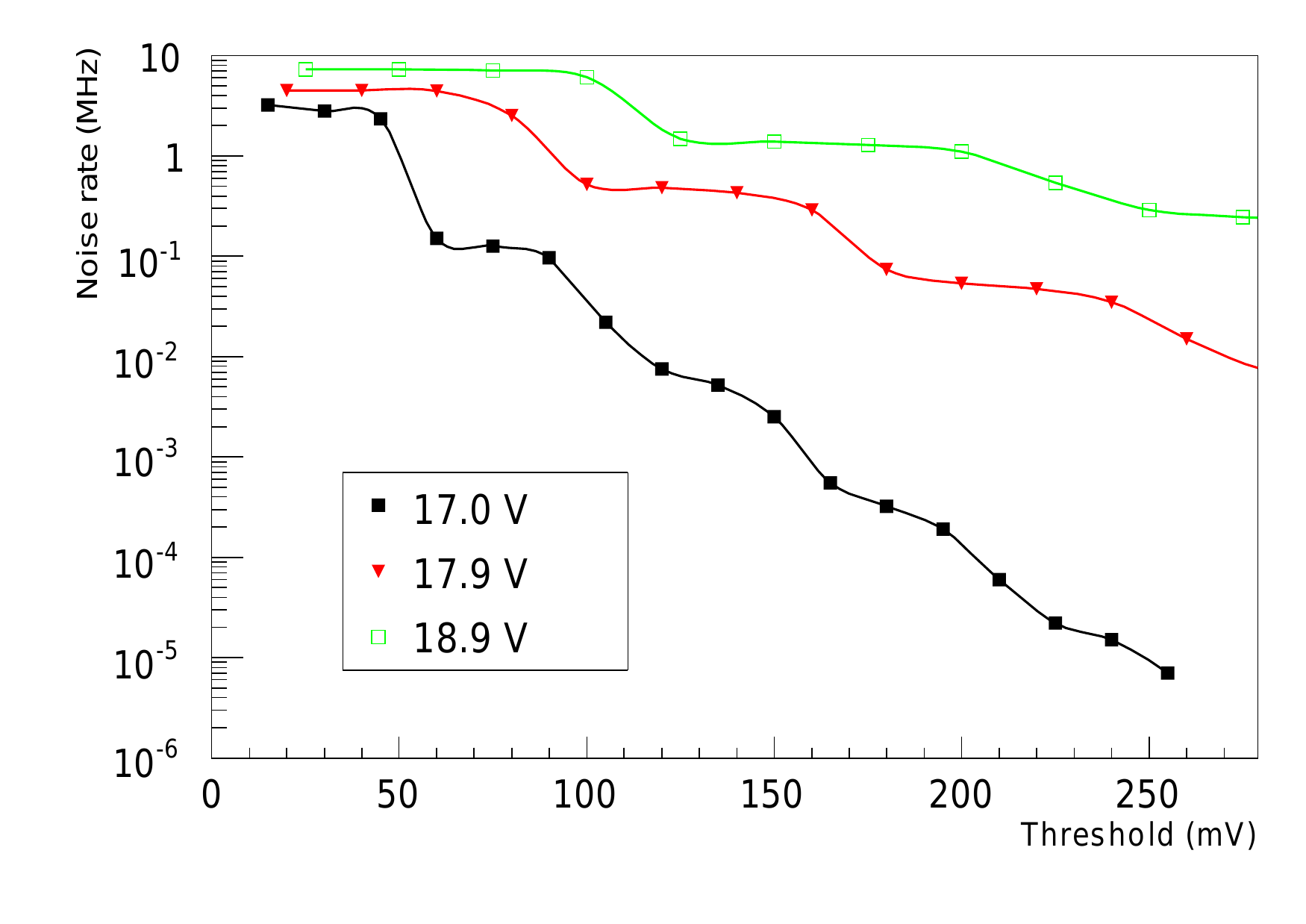}
  \caption{Measured noise count rate as a function of threshold in a
    leading edge discriminator. The plot shows curves for three
    different bias voltages. The noise rate at low thresholds limits
    the use of this SiPM in low light level detection.}
  \label{fig:ratevsthres}
\end{figure}

A critical issue in the operation of SiPM$/$fibre assemblies,
especially when such a detector is self-triggering, is the noise
rate. Two paths are followed in our development of SiPM$/$fibre
tracking systems: Firstly through the increase of the signal
amplitudes by matching the geometries and spectral responses, and
secondly through the use of a versatile trigger selecting the wanted
processes and suppressing the majority of background events.

The detection efficiency is a major issue in designing a tracking
system and the right combination of fibre and SiPM has to be found.
Table~\ref{tab:eff} shows the results of a efficiency measurement for
a 2\,m long fibre read out in both extremes by the 1\,mm$^2$
Photonique device SSPM-0701BG-TO18~\cite{Photonique}. The operating
voltage of the diodes was 17.9\,V.  The fibre was of the standard
(Non-S) type SCSF-78 ({\sf Kuraray}, Japan) with double cladding and
$\oslash=$ 0.83\,mm outer diameter. The emission spectrum extends from
$\lambda\sim$ 415--550\,nm with a maximum at $\lambda\sim$ 440\,nm for
fibres of short lengths. The trapping efficiency for photons produced
close to the axis of the fibre is $5.3\,\%$ giving $70\,\%$ more light
than single cladding fibres, where the efficiency for light trapped
inside the core is only $3.1\,\%$.  This experimental set-up is
labeled with {\em A}.  Minimum ionizing particles are chosen by their
energy deposition in a thick scintillator and coincidences with two
diametrically opposed fibres read-out by conventional photomultipliers
are required for a countable event.  High efficiencies were only
obtained with very low discriminator thresholds.

\begin{figure*}
  \begin{center}
    \includegraphics[height=0.4\textwidth,angle=90]{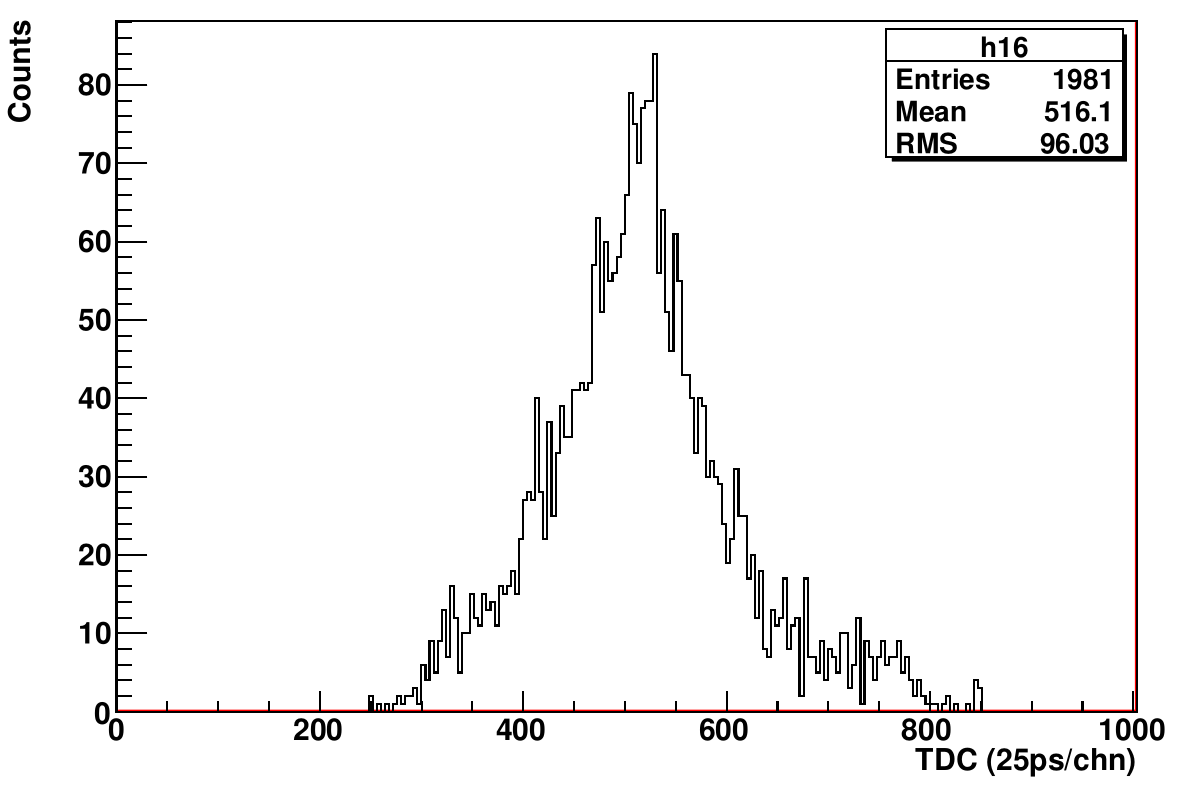}
    \includegraphics[height=0.4\textwidth,angle=90]{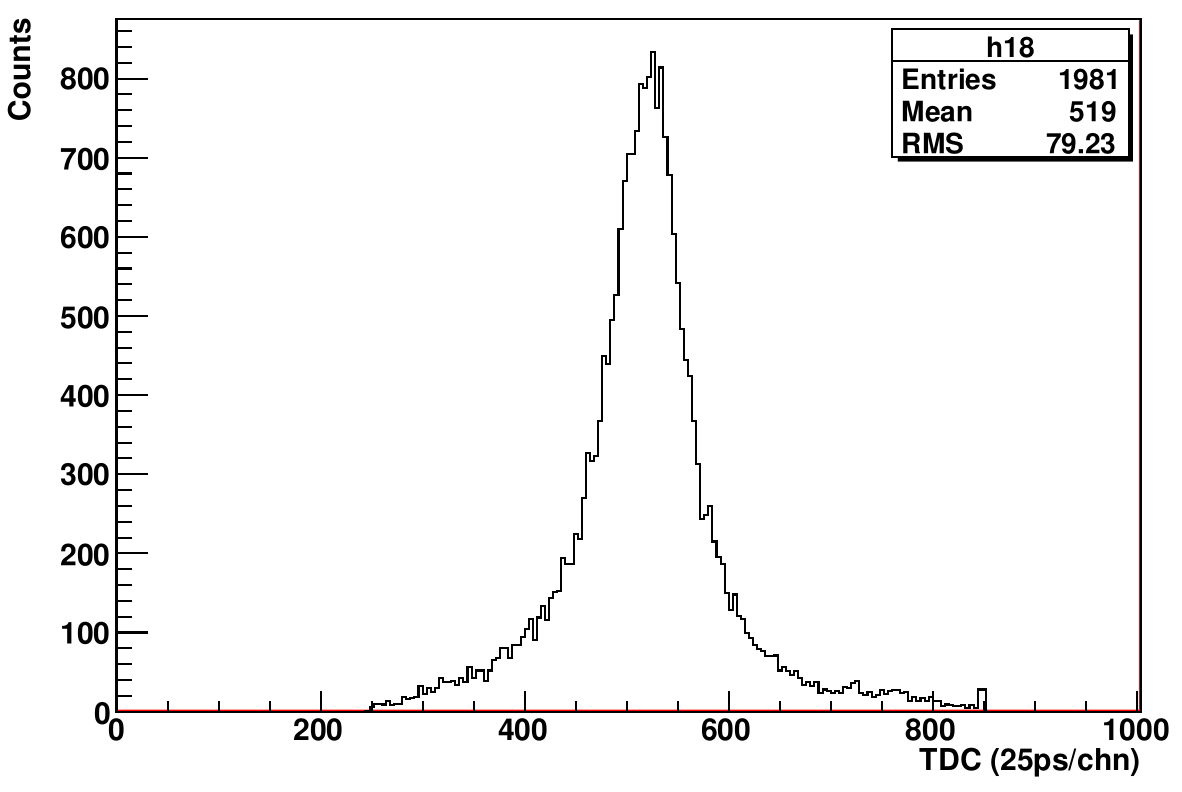}\\
    \includegraphics[height=0.4\textwidth,angle=90]{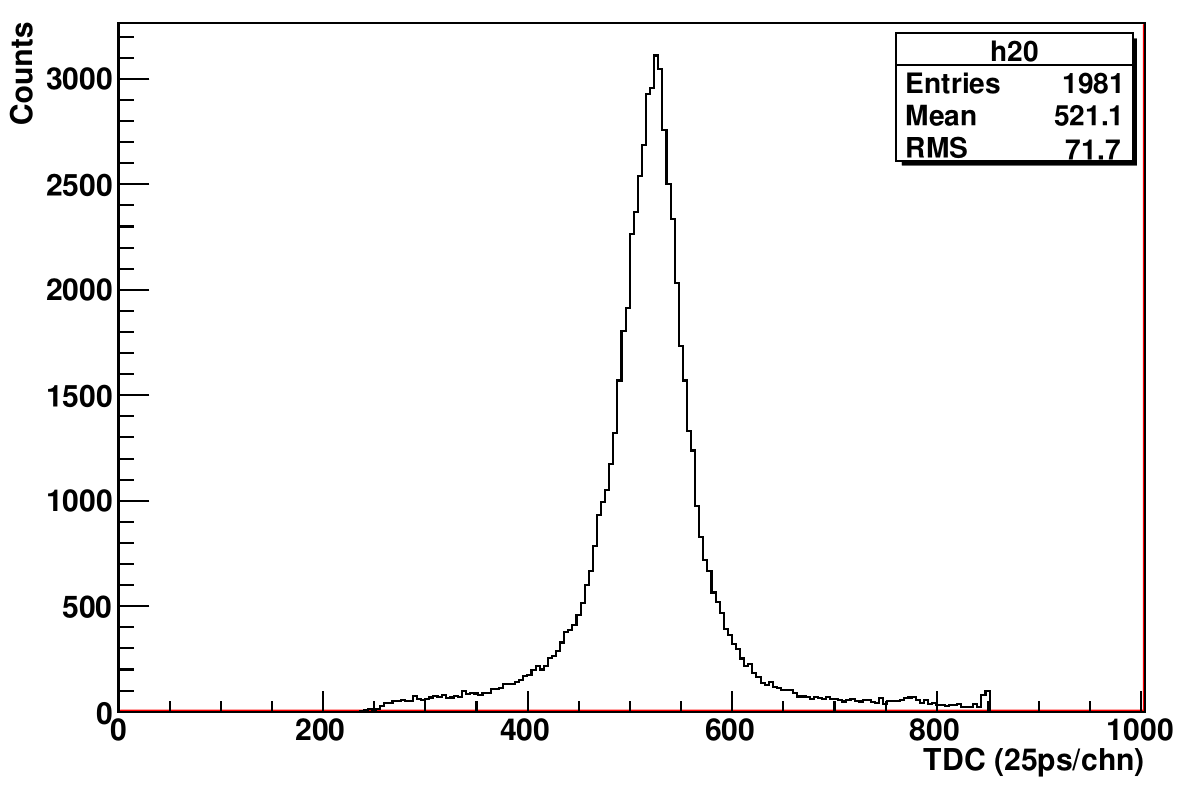}
    \includegraphics[height=0.4\textwidth,angle=90]{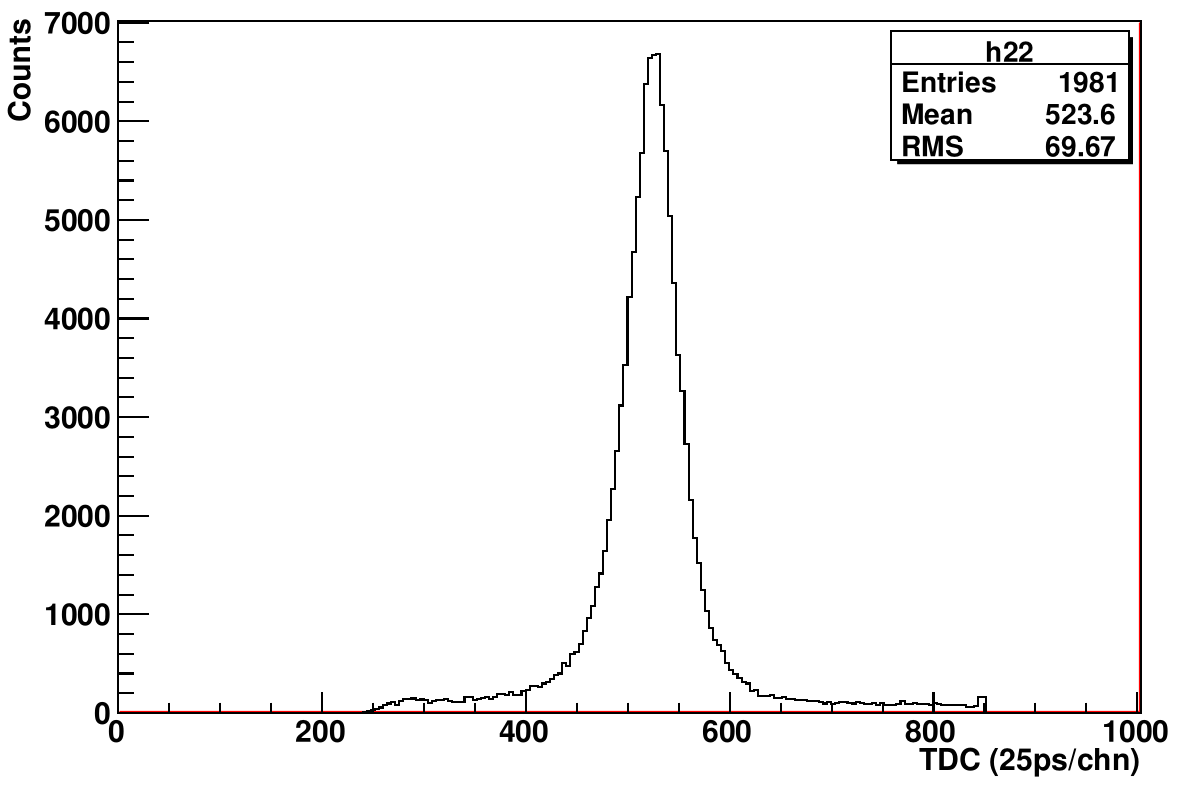}
    \caption{Measured left$/$right time differences with a 2\,m long
      SiPM$/$fibre set-up for signals of $N= $ 1, $\ldots$, 4
      pixels. For the events individual pixels from the ADC spectrum
      were gated, resulting in single detector resolutions of 1.69\,ns
      (rms) for 1 pixel, and 1.40, 1.26, 1.23\,ns (rms) for 2--4
      pixels.}
    \label{fig:SiPMtiming}
  \end{center}
\end{figure*}

A new set-up, which is labeled with {\em B}, has been set-up with
green sensitive SiPM devices and 2\,mm square cross-section green
light emitting fibres of type BCF-20 (emission peak at 492\,nm) with
double cladding from Bicron~\cite{Bicron}. For the SSPM-0606BG4MM-PCB
device typical photon detection efficiencies range from 15 to 25\%
with the value almost constant over the wavelength range 500--650\,nm.
Laboratory measurements with this set-up have shown a signal amplitude
increase by a factor of 2. In 1\,m fibre length a 42\% reduction in
the number of photons was found. The particle detection efficiency was
measured with a modified system of smaller systematic error and the
results are shown in Table~\ref{tab:eff}.  At 3.5\,pixels threshold a
40\,Hz random coincidence rate was observed.  We concluded that a
4\,mm$^{2}$ double cladding fibre with 4\,mm$^2$ SiPM read-out could
be used for the \KAOS\ spectrometer's electron arm tracking detector.

Depending on the bias voltage the SiPM show a noise count rate of up
to 1\,MHz even for thresholds above 1 pixel, see
Fig.~\ref{fig:ratevsthres}. In order to define the hit position in the
fibre detector, a fast clustering algorithm is needed. Coincidences
between left$/$right signals are mandatory. Further, the spectrometer
will be used for experiments with very different experimental
conditions. As a consequence, a versatile trigger system is mandatory
which can be easily reconfigured and adopted to the experimental
requirements. It is obvious that a simple TOF trigger does not fulfill
these requirements. For the electron detector in the \KAOS\
spectrometer a trigger is foreseen, that is based on fast,
programmable multi-purpose logic modules which have been developed
recently at the GSI~\cite{Minami2007}.  Each so called VUPROM module
(VME Universal Processing Module) has 256 channels I$/$O with a high
speed differential LVDS standard. The trigger for the electron arm
will be based on hit multiplicities and hit pattern in the different
fibre planes and will reject background events that do not originate
in the target.

\section{SiPM in PANDA time-of-flight detectors}
At \PANDA\ relatively low momentum $\Xi^-$ can be produced in $\Xi^-
\overline{\Xi}^+$ or $\Xi^- \overline{\Xi}^0$ pair
productions~\cite{PANDA2005}. The associated $\overline{\Xi}$ will
scatter or annihilate inside the residual nucleus. The annihilation
products contain at least two anti-kaons that can be used as a tag for
the reaction. In combination with an active secondary target, high
resolution $\gamma$-ray spectroscopy of double hypernuclei will become
possible for the first time. For the tracking and stopping of the
produced cascade hyperons and their decay products an active
hypernuclear target is needed. For the high resolution spectroscopy of
excited hypernuclear states an efficient, position sensitive germanium
array is required.

In the current \PANDA\ design, particle identification for slow
particles will be provided only at large polar angles by a TOF
detector~\cite{PANDA2005}. As candidates for this detector
scintillator bars and strips or alternatively multi-gap resistive
plate chambers are considered. In the absence of a start detector the
relative timing of multiple particle hits have to be employed. A small
fibre barrel read-out by SiPM has been discussed as an option for a
start detector dedicated to the hypernuclear Physics programme. Such a
detector might be also used as a time reference for the DIRC detector
or for track deconvolution of the time projection chamber.  The SiPM
is intrinsically a very fast detector and its single photoelectron
time resolution is about 100\,ps (FWHM). When coupled to thin and
short scintillating fibres the timing properties are fully dominated
by the scintillation time constants and depend only on the average
number of detected photons. For such a sub-detector system the
achievable time resolution at minimum detector mass is a main
issue. In measurements using the electromagnetic calorimeter the
additional material will deteriorate considerably its resolution.

\begin{figure}[htbp]
  \begin{center}
  \includegraphics[width=0.7\textwidth]{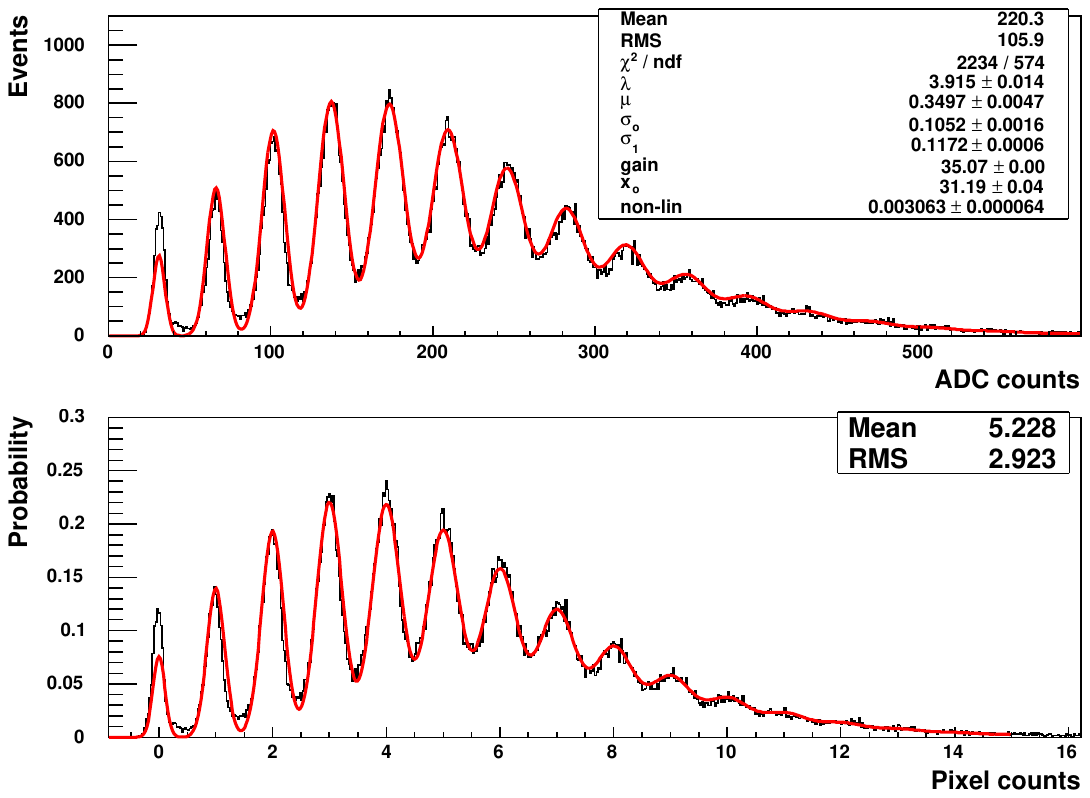}
  \caption{ADC spectrum from a fibre exited by $^{90}$Sr and read out
    by a SiPM. The fit follows a multi-Poissonian distribution taking
    cross-talk into account. The parameters are defined in the text.}
    \label{fig:SiPMspectrum}
  \end{center}
\end{figure}

The time resolution of a single SiPM$/$fibre assembly when excited by
minimum ionizing particles crossing its center was measured by taking
the time difference between the left and right signal simultaneously
to the ADC spectrum. A gate on individual peaks made it possible to
determine the time resolution as a function of the number of fired
pixels as is shown in Fig.~\ref{fig:SiPMtiming}.  With the set-up
discussed in Section~2 single detector SiPM resolutions of 1.69\,ns
(rms) for 1 pixel, and 1.40, 1.26, 1.23\,ns (rms) for 2--4 pixels were
determined. The amplifier was type AMP\_06011 from Photonique with a
rise-time of 700\,ps and signal amplification of 10--20 times. The
time distributions are non-Gaussian and are well described with
exponential functions on both sides of the peak. The structure at
around TDC channel 750 was created by a reflection of photons on the
SiPM surface and their detection by the opposite SiPM in 2\,m
distance.  To determine the photoelectron yield from the ADC spectrum
as shown in Fig.~\ref{fig:SiPMspectrum} a multi-Poissonian fit of the
following type was performed:
\begin{eqnarray*}
  f(q) & = & \sum_{p=0}^{15} \sum_{s=0}^{15}
    \left( \frac{\exp^{-\lambda}\lambda^p}{p!} \right)
    \left( \frac{\exp^{-p\mu}(p\mu)^s}{s!} \right) \\
    & & \times \left( \frac{
        \exp{\left(-\frac{1}{2}\frac{[q-(p+s)]^2}{
            \sigma_0^2+(p+s)\sigma_1^2} \right) } }{
        \sqrt{2\pi}\sqrt{\sigma_0^2+(p+s)\sigma_1^2}} \right)
\end{eqnarray*}
with $q= \frac{x-x_0}{g} - nl \cdot \left( \frac{x-x_0}{g}\right)^2$,
$x, x_0=$ ADC counts, pedestal off-set, 
$p, \lambda=$ photon counts and mean value,
$s, \mu=$ secondary pixel counts and mean value, and
$\sigma_0, \sigma_1=$ pedestal and pixel noise.
By using this fitting method, which takes optical cross-talk into
account, mean numbers of photoelectrons $\overline{N}\sim$ 4 were
measured with a $^{90}$Sr source. A Monte Carlo model for photon
generation and tracking was applied to model a detector with 10\,cm
long double cladding fibres verifying the measurements for a possible
\PANDA\ detector. Based on the simulation a time resolution of
0.9\,ns\,(rms) is expected.

\begin{figure}[htbp]
  \begin{center}
    \includegraphics[width=0.7\textwidth]{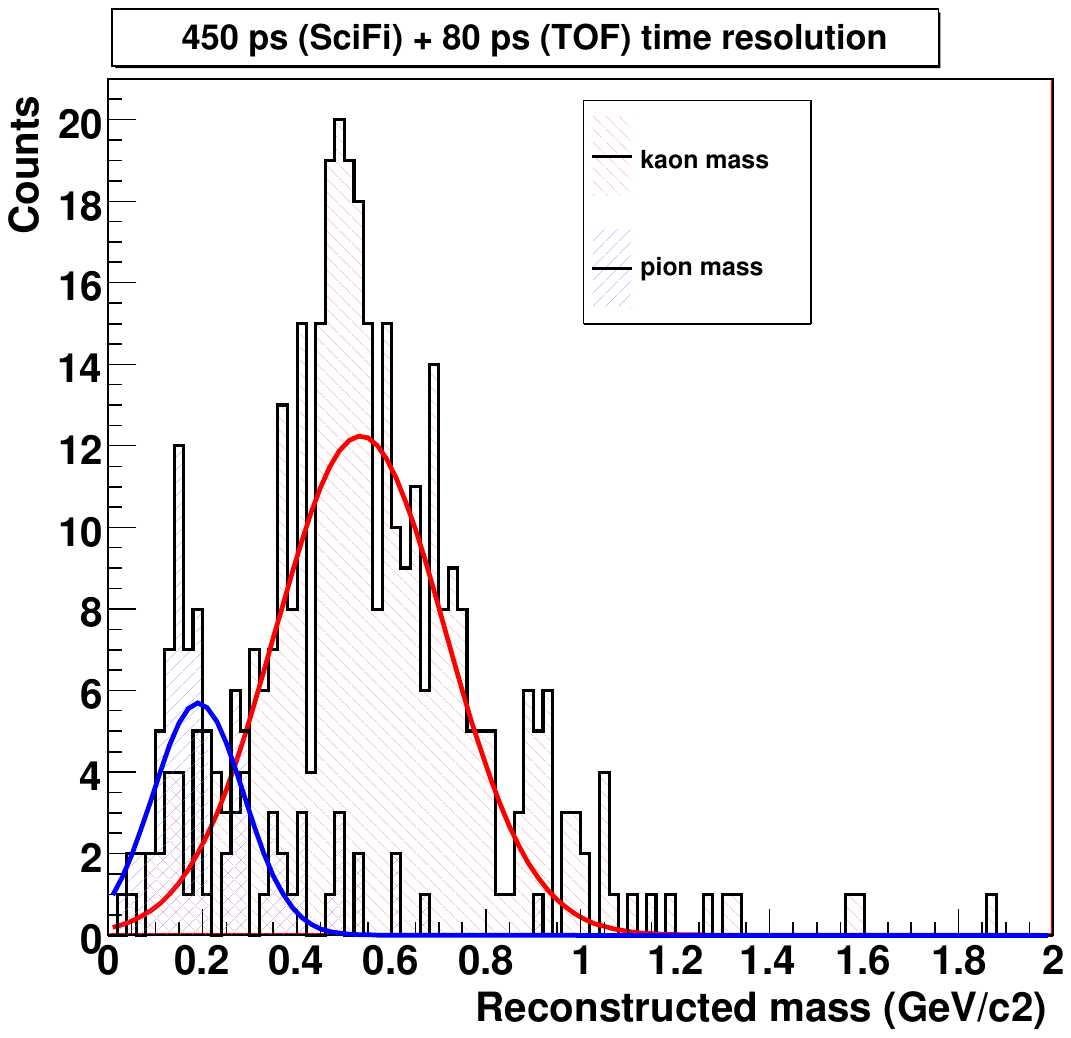}
    \caption{Simulated low momentum kaon$/$pion separation for the
      \PANDA\ hypernuclei programme using a fibre start counter and a
      barrel stop counter of scintillating bars and strips with 80\,ps
      and 450\,ps time resolution, respectively. Calculated with 1\%
      momentum resolution and 2\% error in track length.}
    \label{fig:PANDA_TOF}
  \end{center}
\end{figure}

A full GEANT4 simulation of the $\gamma$-ray spectroscopy set-up with
several germanium cluster detectors has been
completed~\cite{Lorente2008}. The simulation was performed using a
demanding time-of-flight (TOF) detector system for low momentum kaon
identification: a start detector of $\sim$ 2000 scintillating fibres
placed in two rings and a TOF barrel detector of 16 slabs. As can be
deduced from the results seen in Fig.~\ref{fig:PANDA_TOF} the fibre
detector has to provide the start time with a minimum resolution of
$\sigma\sim$ 400\,ps, if one assumes a time resolution of $\sigma <$
100\,ps for the stop detector. It seems challenging to achieve these
values with the geometries and photon detection devices described in
this paper.

\section{Discussion}
SiPM technology is evolving very fast and one of the main performance
improvements is the progressive reduction of the dark count rate. This
new situation makes scintillating fibre tracking detectors read-out by
SiPM an interesting and realistic option.  Fibres with larger ($\sim$
2\,mm) cross-sections were generating enough light to get an
acceptable detection efficiency for higher thresholds reducing
considerably the noise level. On the other hand, larger SiPM do
produce higher dark count rates.

Low light level detection will be degraded or even impossible if the
devices show an increasing rate of dark pulses when operated in an
radiation environment. The effects of irradiation on the
characteristic parameters of SiPM such as dark count rate or gain
uniformity were determined and found to be
relevant~\cite{Sanchez2008}. A dedicated study of hadronic radiation
effects on SiPM in the \PANDA\ environment is necessary before a final
conclusion can be drawn.

Our study has shown that present devices can be used to realise a TOF
detector with the required properties. To fully exploit the good
timing properties of SiPM, intensive development has to be carried out
to increase the output signal of SiPM$/$fibre assemblies well above
the noise level.

\section*{Acknowledgements}
Work supported in part by the European Community under the
``Structuring the European Research Area'' Specific Programme as
Design Study DIRAC\-secondary-Beams (contract number 515873).



\end{document}